\documentclass[conference,10pt]{IEEEtran}
\IEEEoverridecommandlockouts
\usepackage{cite}
\usepackage{amsmath,amssymb,amsfonts}
\usepackage{algorithm,algorithmic}
\usepackage{graphicx}
\usepackage{textcomp}
\usepackage{color,xcolor}
\usepackage{bm}
\usepackage{multirow}
\usepackage[normalem]{ulem}
\usepackage{tabularx}
\usepackage{cellspace}
\usepackage{xpatch}
\usepackage{soul}
\usepackage{ulem}
\usepackage{verbatim}
\usepackage{caption}

\usepackage{subcaption}
\usepackage[top=0.65in, bottom=1.22in, left=0.61in, right=0.62in]{geometry}

\begin{document}

\title{Wavenumber Domain Sparse Channel Estimation \\
	 in Holographic MIMO


\thanks{Xufeng Guo and Yuanbin Chen contributed equally to this work.

% (\textit{Corresponding author}: Ying Wang.)
}
}
 
\author{{\large Xufeng~Guo\textsuperscript{1,*},~Yuanbin~Chen\textsuperscript{1,*},~Ying~Wang\textsuperscript{1},~Zhaocheng~Wang\textsuperscript{2},~and~Zhu Han\textsuperscript{3}}\\
	{\normalsize \textsuperscript{1}State Key Laboratory of Networking and Switching Technology,}\\ {\normalsize Beijing University of Posts and Telecommunications, Beijing 100876, China}\\
	%	{\normalsize \textsuperscript{1}Beijing University of Posts and Telecommunications, Beijing 100876, China}\\
	{\normalsize \textsuperscript{2}Beijing National Research Center for Information Science and Technology,} \\
 {\normalsize Department of Electronic Engineering, Tsinghua University, Beijing 100084, China}\\
	% 	{\normalsize \textsuperscript{2}Beijing Institute of Technology, Beijing 100081, China}\\
	{\normalsize \textsuperscript{3}Department of Electrical and Computer Engineering, University of Houston, Houston, TX 77004 USA}
	% {\normalsize \textsuperscript{4}...} \\
	%	{\small   \textsuperscript{*}Zixing Tang and Yuanbin Chen contributed equally to this work.  }
	
	%	Email: wangying@bupt.edu.cn, maotq@buaa.edu.cn, qingqingwu@sjtu.edu.cn,lh@ecs.soton.ac.uk, marco.di-renzo@universite-paris-saclay.fr
%	\vspace{-5mm}
}

\maketitle

\begin{abstract}

In this paper, we investigate the sparse channel estimation in holographic multiple-input multiple-output (HMIMO) systems. The conventional angular-domain representation fails to capture the continuous angular power spectrum characterized by the spatially-stationary electromagnetic random field, thus leading to the ambiguous detection of the significant angular power, which is referred to as the power leakage. To tackle this challenge, the HMIMO channel is represented in the wavenumber domain for exploring its cluster-dominated sparsity. Specifically, a finite set of Fourier harmonics {\color{black}acts} as a series of sampling probes to encapsulate the integral of the power spectrum over specific angular regions. This technique effectively eliminates power leakage resulting from power mismatches induced by the use of discrete angular-domain probes. Next, the channel estimation problem is recast as a sparse recovery of the significant angular power spectrum over the continuous integration region. We then propose an accompanying graph-cut-based swap expansion (GCSE) algorithm to extract beneficial sparsity inherent in HMIMO channels. Numerical results demonstrate that this wavenumber-domain-based GCSE approach achieves robust performance with rapid convergence.

\end{abstract}


\section{Introduction}

Holographic multiple-input multiple-output (HMIMO) is envisioned as a prospective and potential technology that holds promise for meeting the high demands of the sixth generation (6G) communications, poised to realize holographic radio with reasonable power consumption and fabrication costs \cite{Holo-21, Holo-A-105}. In contrast to traditional antenna arrays based on discrete antenna elements, the HMIMO approach boasts an almost continuous antenna surface, capable of generating any current distribution to fully exploit the propagation characteristics of electromagnetic (EM) channels. 
The application of densely arranged antenna elements in HMIMO enables a nearly continuous aperture, resulting in a large improvement in spatial resolution and the ability to achieve super-directivity. This, in turn, leads to remarkable enhancements in beamforming capabilities \cite{A-2,A-3}.

To fully unlock the potentials of HMIMO systems, establishing its propagation channel model and efficiently acquiring channel state information (CSI) is of paramount importance, serving as the bedrock for further efficient transmissions. Regarding the HMIMO channel model, there are some existing works~\cite{28.6,Holo-A-105,Holo-21,GC-tengjiao} in the literature. 
These studies converge on a consensus, modeling the HMIMO channel as a spatially-stationary EM random field. In particular, an exact statistical representation is given by the Fourier plane-wave series expansion. This leads to a stochastic description of the EM channel where the array geometry and scattering can be separated. Based on this class of channels, numerous efforts have been devoted to spectral efficiency analysis \cite{A-4}, precoding design \cite{Holo-14}, and channel estimation \cite{Holo-4}. 

For the HMIMO channel estimation literature, a channel estimation scheme based on the least square (LS) estimator is proposed in \cite{Holo-4}. However, due to the intrinsic nature of the LS estimator, matrix inversion operations are inevitable. In the presence of high-dimensional HMIMO channel induced by numerous antenna elements, conventional channel estimating techniques such as LS and minimal mean square error (MMSE) can be challenging to implement and entail a significant amount of pilot overhead. 
%In addition, it should be noted that the process of matrix inversion becomes problematic when dealing with channel matrices of exceptionally high dimensions. This issue, commonly referred to as a ``dimensional disaster", poses significant challenges that make the actual implementation of such operations unfeasible.
Inspired by the sparsity observed in massive MIMO (mMIMO) channels, where a limited number of scatterers in the environment lead to a few significant paths, the angular domain channel in mMIMO exhibits beneficial sparsity. The sparse estimation in the angular domain can achieve pilot savings, scaling it down from a magnitude comparable to the number of antennas to a level proportionate to the number of significant paths \cite{34.,12.}. 
Therefore, the above literature review evokes a reminiscent inquiry: {\em Can there also be angular domain sparsity in HMIMO channels}? This speculation is further fueled by the channel model in \cite{28.6}, which includes a stochastic description of clusters in the environment, greatly sparking our interest and motivation for this work.

However, the angular domain representation does not perfectly extract the sparsity inherent in HMIMO channels. Specifically, the traditional discrete Fourier transform (DFT)-based sparsifying basis essentially samples the Dirichlet kernel discretely in the angular domain, but this approach is only suitable for configurations with half-wavelength antenna spacing. In HMIMO scenarios, where the antenna spacing is far less than half a wavelength, the mismatch between the sampling probes and the zero points of the Dirichlet kernel leads to ambiguous detection of significant angular power, which we refer to as power leakage. Therefore, to explore the potential sparsity of HMIMO channels, we shift our focus to the wavenumber-domain representation as specified in \cite{28.6}. Explicitly, the wavenumber-domain representation is based on a series of Fourier harmonics (FHs), which is in essence a series of orthogonal basis functions that can be used to represent the integral of the power spectrum over a corresponding angular region for a continuous or near-continuous aperture array. This approach can mitigate the power leakage issues present in the angular domain and facilitate more precise detection of cluster-dominated HMIMO channel sparsity.

Although the concept of the wavenumber domain has been proposed, to our knowledge, there is a significant gap in the literature in fully exploring the sparsity of HMIMO channels within this domain. Thus, this work takes the first attempt at wavenumber-domain sparse channel estimation for HMIMO, as it addresses the non-trivial challenges posed by the substantial number of antenna elements involved in HMIMO channel estimation.
Geared towards above-mentioned challenges, this work aims for filling the knowledge gap in the state-of-the-art by following contributions:
\begin{itemize}
	\item We first demonstrate the adverse power leakage of the HMIMO channel in the angular domain. Specifically, the significant angular power spectrum dominated by the clusters in propagation environment cannot be clearly identified in the angular domain. Accordingly, a novel dictionary matrix is crafted based on Fourier harmonics sparsifying basis to restructure the original HMIMO channel as a sparse one.
	
	\item  The sparse HMIMO channel estimation is formulated as a compressed sensing (CS) problem. In contrast to conventional angle-wise estimation in the angular domain, the individual non-zero entry in the sparse vector fails to adequately encapsulate the characteristics of single cluster present in the environment. Hence, these channel pieces need to be spliced together within the continuous integration region in the wavenumber domain for achieving a complete clustering description.
	
	\item  To facilitate robust recovery of significant angular power spectrum in HMIMO channels, Markov random field (EMRF) model is employed to capture the correlations among different non-zero entries. Then a graph-cut-based swap expansion (GCSE) algorithm is proposed to estimate the wavenumber-domain channel. Simulation results are provided to verify the effectiveness of the proposed approach.

\end{itemize}

\section{System Model}
In an uplink HMIMO communication system, we consider a base station (BS) equipped with a uniform planar array (UPA) serving a single-antenna user.
%{\footnote{\color{red}This can be readily extended to multi-user case, and in this work we just unveil the structured sparsity in the wavenumber domain}}
The UPA comprises  $N = N_x \times N_y$ antenna elements, where $N_x$ and $N_y$ denote the number of antenna elements along the $x-$axis and $y-$ axis, respectively. The horizontal length of the UPA is $L_x = (N_x-1)\delta$ while the vertical is $L_y = (N_y-1)\delta$, where $\delta$ represents the antenna spacing being far below half of the wavelength $\lambda$, i.e., $\delta \ll \lambda /2 $. There are $N_{f}$ feeds positioned on the UPA for generating reference waves that carry user-intended signals, with each feed attached to a radio frequency (RF) chain for signal processing. The uplink signal $\mathbf{y} \in {\mathbb{C}}^{N_{f} \times1}$ received at the BS is given by
\begin{equation}\label{eq:transmit_model}
 \mathbf{y} = \text{diag} \left( \mathbf{A}\right)  \mathbf{P} \text{diag} \left( \mathbf{M}\right)  \mathbf{H} x + \mathbf{n} = \mathbf{C} \mathbf{H} x + \mathbf{n},
\end{equation}
where ${\bf C} = \text{diag} \left( \mathbf{A}\right)  \mathbf{P} \text{diag} \left( \mathbf{M}\right) $ is the composite combining matrix with $\mathbf{A} \in {\mathbb{C}}^{N_{f} \times 1}$, $\mathbf{M} \in {\mathbb{R}}^{N \times 1}$ and $\mathbf{P} \in {\mathbb{C}}^{N_{f} \times N}$ being the {digital precoding vector}, the amplitude-controlled holographic beampattern matrix, and the phase difference matrix, respectively. $\mathbf{n} \in {\mathbb{C}}^{N_{f} \times 1}$ is the additive white Gaussian noise (AWGN) vector with each entry following $\mathcal{CN}(0,\sigma_{\rm noi}^2)$.
% is the digital precoding vector and  denotes the amplitude-controlled holographic beampattern matrix. Matrix $\mathbf{P} \in {\mathbb{C}}^{N_{f} \times N}$ encapsulates the phase differences that the EM wave undergoes as it propagates from the feed to each antenna element, whose $\left( {n_f, n} \right) $-th entry is denoted by $ {{{\left[ {\mathbf{P}} \right]}_{{n_{f}},n}}}  = \exp{\left\{jk_c r(n_f, n) \right\}}$, with $k_c$ representing the wavenumber of the carrier frequency and $r(n_f,n)$ for the distance between the $n_f$-th feed and the $n$-th antenna element. 
The signal $x$ transmitted by the user satisfies $| x | = 1$. 
% For simplicity, we take the unit power assumption, i.e., $x = 1$.
%{\color{red}For simplicity, we only consider the channel in a single snapshot, i.e., $x = 1$.} 
The HMIMO channel $\mathbf{H}$ fundamentally follows the Fourier planar-wave series expansion-based channel model presented in \cite{28.6,Holo-21,GC-tengjiao}, 
% which in essence captures the spatially-stationary small-scale fading characteristics. 
in this case, $\mathbf{H}$ is given by
\begin{equation}\label{HMIMO_channel}
	{\bf H} = \sum_{l\triangleq \left(l_x, l_y\right)\in \mathcal{L}} h_l^f {\bf a}^f \left( l_x, l_y \right) .
\end{equation}
In (\ref{HMIMO_channel}), $ {\bf a}^f \left( l_x, l_y \right)$ is an FH-based steering vector with the $n\triangleq (n_x,n_y)$-th entry ($n_x$, $n_y$ are the horizontal and vertical antenna indices, respectively) denoted by 
\begin{equation}\label{FH_steering}
	\left[{\bf a}^f \left( l_x, l_y \right)\right]_n = \exp \left\lbrace {j\left(  \frac{2\pi l_x}{L_x}\delta n_x + \frac{2\pi l_y}{L_y}\delta n_y \right) } \right\rbrace ,
\end{equation}
in which $l\triangleq \left(l_x, l_y\right) \in \mathbb{Z}^2$ denotes the two-dimensional (2D) index of the FHs, with $l_x$ and $l_y$ representing the associated horizontal and vertical indices, respectively. $\mathcal{L}$ is a collection of indices associated with FH basis, given by 
\begin{align}\label{eq:xi_defination}
%		\mathcal{L} \triangleq\left\{\left(l_x, l_y\right)\in \mathbb{Z}^2 \mid\left(\frac{2 \pi }{L_x}l_x\right)^2+\left(\frac{2 \pi }{L_y}l_y\right)^2 \leq k_c^2\right\} \nonumber \\
 \mathcal{L} \triangleq\Bigg\{\left(l_x, l_y\right) \in \mathbb{Z}^2\mid\left(\frac{\lambda}{L_x} l_x \right)^2  +\left(\frac{\lambda}{L_y} l_y \right)^2  \leq 1\Bigg\}.
\end{align}
The FH basis collection $\mathcal{L}$ can also be understood as a unit sampling collection of an ellipse region in the wavenumber domain, the area of which is $\pi \frac{L_x L_y}{\lambda^2}$, thus, $L\triangleq |\mathcal{L}| \approx \lfloor \pi \frac{L_x L_y}{\lambda^2} \rfloor$.

Still referring to (\ref{HMIMO_channel}), $h_l^f$  captures the HMIMO channel characteristics, and follows a complex Gaussian distribution with zero mean and variance {\small $\left( \sigma_l^f\right)^2$} as following
\begin{subequations}\label{h_l_f}
	\begin{align}
		h_l^f &\sim \mathcal{C N}\left(0,\left(\sigma_l^f\right)^2\right), \forall l \in \mathcal{L}, \\
		\left(\sigma_l^f\right)^2&=\iint_{\Omega_f\left(l_x, l_y\right)} A^2(\theta, \phi) \sin \theta \mathrm{d} \theta \mathrm{d} \phi ,
	\end{align}
\end{subequations}
where $\Omega_f\left(l_x, l_y\right)$ is the angular integration region corresponding to the $l$-th wavenumber domain~\cite[Appendix]{28.6.8}, and $A^2(\theta, \phi)$ represents the angular power spectrum in propagation environments, carrying the clustered characteristics and serves as a function of both the elevation angle $\theta$ and the azimuth angle $\phi$, given by
\begin{equation}\label{eq_A}
	A^2\left(\theta, \phi\right) = \sum_{i\in \left\{ 1,\dots, N_c \right\}} w_i p_i \left( \theta, \phi \right).
\end{equation}
In (\ref{eq_A}), $w_i$ is the weight of the $i$-th cluster ($N_c$ clusters in total), satisfying $\sum_{i\in \left\{ 1,\dots, N_c \right\}} w_i = 1$, $\alpha_i$ is the concentration parameter of the $i$-th cluster, and $\left( \theta_i, \phi_i \right)$ denote angular spectrum center of the $i$-th cluster. $p_i \left( \theta, \phi \right)$ represents  the probability distribution function of the three-dimensional (3D) von Mises-Fisher (VMF) distribution \cite{28.6}, i.e.,
\begin{equation}\label{VMF_probability}
	\resizebox{1.0\hsize}{!}{$
		p_i\left( \theta, \phi \right)  = \frac{\alpha_i}{4\pi \sinh \alpha_i} 
		\exp  \left\lbrace {\alpha_i \left\{
			\sin\theta \sin\theta_i \cos\left( \phi - \phi_i \right) + \cos\theta\cos\theta_i
			\right\} } \right\rbrace .$}
\end{equation}

Based on the channel representations shown in (\ref{HMIMO_channel}) and (\ref{eq_A}), one can intuitively infer that the sparsity of HMIMO channels might be dependent on the clusters present in the propagation environment. This understanding might naturally lead us to leverage the traditional DFT-based angular-domain representation as a plausible method to examine the sparsity of HMIMO channels. However, in the angular domain, it is challenging to identify the angular power spectrum specific to each cluster. Thus, in the section that follows, we aim to verify the infeasibility of observing sparsities of HMIMO channels in the angular domain, while unveiling their significant sparsities in the wavenumber domain.

% \begin{figure}[t]
% %	\centerline	
% \includegraphics[width=0.98\linewidth]{figure/AD_WD_compare/compare1.pdf}
% 	\caption{Angular power indication of the propagation channel in the (a) angular domain, and (b) wavenumber domain.}
% \label{fig:AD_power_tail}
% 	%	\vspace{-5mm}
% \end{figure}

% \begin{figure}[t]
% 	%	\centerline	
% 	\includegraphics[width=0.99\linewidth]{figure/AD_compare5.pdf} 
% 	\caption{Angular power `leakage' or `spreading'.}
% 	\label{fig:AD_leakage_reason}
% 	%	\vspace{-5mm}
% \end{figure}

\begin{figure*}[htbp]
  \centering
  \begin{minipage}{0.545\textwidth}
      \centering
      \includegraphics[width=\linewidth]{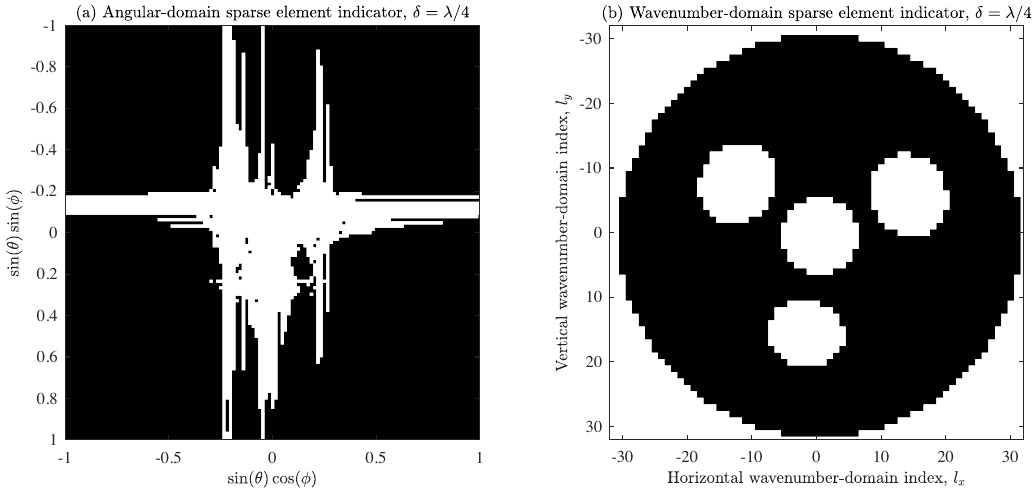}
      \caption{Angular power indication of the propagation channel in the (a) angular domain, and (b) wavenumber domain.}
      \label{fig:AD_power_tail}
  \end{minipage}
  \hfill
  \begin{minipage}{0.42\textwidth}
      \centering
      \includegraphics[width=\linewidth]{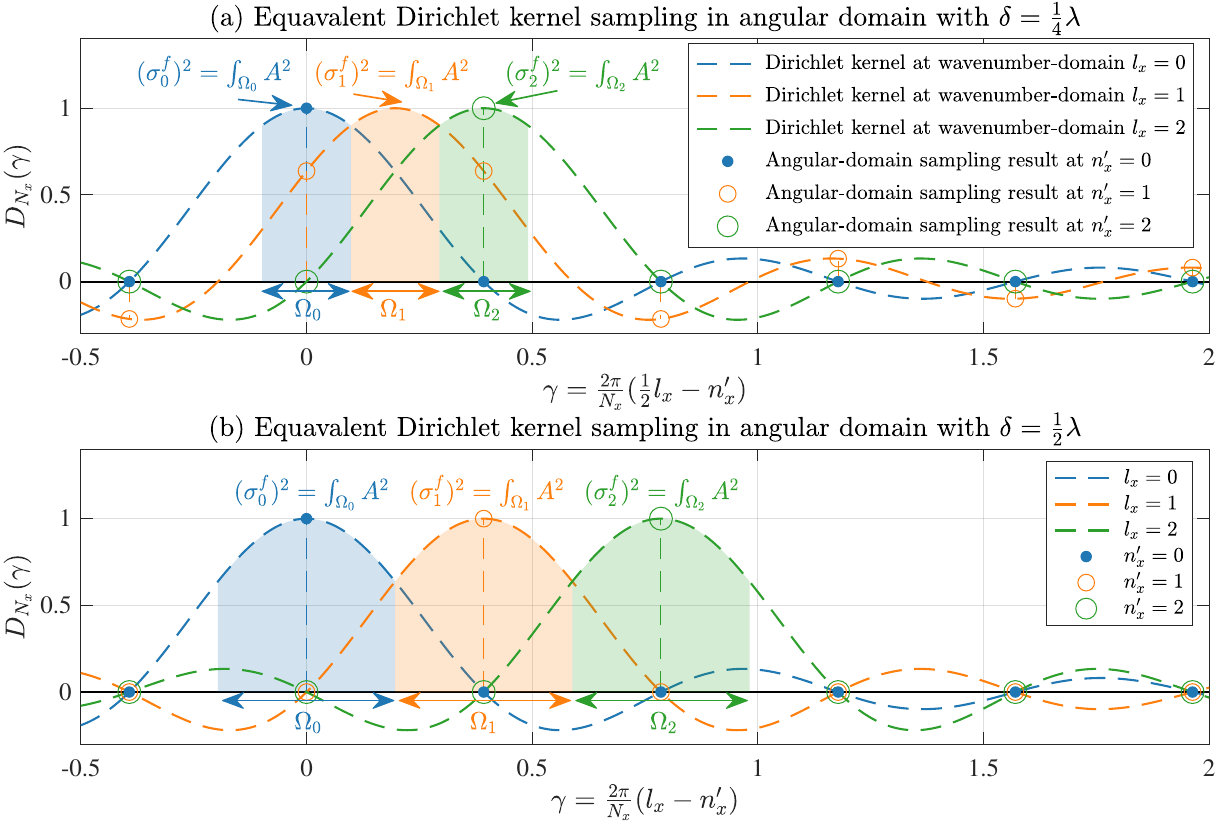}
      \caption{Angular power `leakage' or `spreading'.}
      \label{fig:AD_leakage_reason}
  \end{minipage}
  \vspace{-0.5cm}
\end{figure*}

\section{Sparse Wavenumber Domain Representation}

In this section, we commence by unveiling the challenges of implementing the conventional angular-domain representation in HMIMO systems. Then, a custom-designed wavenumber-domain technique is introduced in Section~\ref{sec:WD_tech} to effectively unmask the sparsity inherent to HMIMO channels.

\subsection{Challenges of Angular Domain Representation}
We denote ${{\mathbf{\Psi }}^a} \in {\mathbb{C}^{N \times N}}$ by the angular-domain dictionary matrix, whose $(n,n^\prime)$-th entry is given by 
\begin{equation}\label{AD_dic}
	{\left[ {{{\mathbf{\Psi }}^a}} \right]_{n,n'}} = \frac{1}{{\sqrt N }}\exp \left\{ {j\left( {\frac{{2\pi {n_x}}}{{{N_x}}}{n^\prime_x} + \frac{{2\pi {n_y}}}{{{N_y}}}{n^\prime_y}} \right)} \right\}, 
\end{equation}
with $\forall n = \left( {{n_x},{n_y}} \right)$ being the index of the antenna element and $n^\prime = \left( {n^\prime_x},{{n^\prime_y}} \right)$ for the sampling points in the angular domain~\cite{Spectral_CS}. Therefore, the HMIMO channel $\mathbf{H}$ can be structured as its angular-domain counterpart, i.e., ${{\mathbf{\Psi }}^a}{{\mathbf{h}}^a}$. In Fig.~\ref{fig:AD_power_tail}(a), we demonstrate the angular power indicators of the angular-domain channel ${\mathbf{h}}^a$, in which $N_c=4$ clusters are configured in the propagation environment. As it transpires, a concentrated white region at the center may indicate the presence of the cluster's angular power. However, the concentrated angular power seems to disperse outward in a manner reminiscent of `leakage' or `spreading', which prevents us from observing the precise clustered sparsity present in the HMIMO channel.

% significant angular power
% angular power spectrum
% mapping mismathch of significant angular power from angular domain to the wavenumber-domain

{\color{black}
This occurrence can be attributed to the mismatch in mapping significant angular power
from the actual angular power spectrum caused by the FH-based steering vector in HMIMO channels to the DFT-based angular domain.
% from the angular domain to the wavenumber domain. 
As illustrated in Fig.~\ref{fig:AD_leakage_reason}, we analyze this mapping through three sampling cases: $n_x^\prime= l_x=0$, $n_x^\prime=l_x=1$, and $n_x^\prime= l_x=2$, to examine their alignment and mismatches. More explicitly, sampling in the angular domain in essence constitutes a few discrete points (impulse responses), while in the wavenumber domain, a continuous Dirichlet kernel is employed to represent the sampling result. In the case of antenna spacing being $\delta = \lambda / 2$, we observe that both the non-zero and zero impulse responses in the angular domain coincide with the peaks and zero points of the Dirichlet kernel in the wavenumber domain. The function $D_{N_x}(\gamma)$ can be employed to mathematically  describe the relationship between $n_x^\prime$ and $l_x$, where $\gamma {\rm =} \frac{2\pi}{N_x} (\frac{2\delta}{\lambda}l_x - n_x^\prime)$~\cite{Spectral_CS}. Thus, when $\delta = \lambda / 2$, significant angular power can be accurately identified, irrespective of its sparsifying basis. By contrast, in the case of $\delta = \lambda / 4$, the sampling $n_x^\prime =1$ fails to detect the true significant angular power represented by $l_x =1$. This indicates that even if the power in the wavenumber domain is not notable, it may be misinterpreted as significant in the angular domain.  Unfortunately, this mismatch tends to occur whenever $l_x$ is an odd number is exacerbated as  antenna density increases (i.e., with smaller $\delta$). For instance, a sampling error probability of $1-2\delta/\lambda=1/2$ arises in the event of $\delta {\rm =}  \lambda /4$, and this probability increases to 3/4 as $\delta$ further decreases to $\lambda / 8$.
}

\subsection{FH-Based Wavenumber-Domain Representation}\label{sec:WD_tech}

Our goal is to craft a dictionary matrix, like a bespoke `filter', for unmasking the inherent sparse characteristics of the HMIMO channel. The wavenumber-domain representation is explicitly based on the Fourier harmonics, which is in essence a series of orthogonal basis functions that can be used to represent the array response for a continuous or near-continuous aperture array~\cite{28.6}. Inspired by this, we restructure the dictionary matrix ${{\mathbf{\Psi }}^f} \in {\mathbb{C}^{N \times L}}$ by harnessing FH basis, whose $\left( {n,l}\right)  $-th entry is given by
\begin{equation}
{\left[ {{{\mathbf{\Psi }}^f}} \right]_{n,l}} = \frac{1}{{\sqrt N }}\exp \left\{ {j\left( {\frac{{2\pi {n_x}}}{{{L_x}}}\delta {l_x} + \frac{{2\pi {n_y}}}{{{L_y}}}\delta {l_y}} \right)} \right\}.
\end{equation}
Then, the HMIMO channel can be recast as ${\mathbf{H}} = {{\mathbf{\Psi }}^f}{{\mathbf{h}}^f}$, and ${{\mathbf{h}}^f} \in {\mathbb{C}^{L \times 1}}$ is the wavenumber-domain channel with only $K$ non-zero entries thereof.  $K$ can be understood as the number of the significant FH basis, in contrast to the  total number of FH basis $L \approx \pi \frac{L_x L_y}{\lambda^2}\gg K$. Therefore, ${\mathbf{h}}^f$ is a $K$-sparse vector with $L$ size.

% \hl{Further explanations (with power tail simulation) understanding of Eq. (9)
% This also requires the explanation of Fig.1(b) with more comprehending understanding
% }
{\color{black}
% In both angular-domain and wavenumber-domain compressive sensing (CS) methods, the output channel is conceptualized as the weighted sum of a certain number of selected columns in the dictionary matrix. These weight factors are the non-zero elements in sparse channel representations, namely ${\bf h}^a$ and ${\bf h}^f$. However, 
With angular-domain modeling, due to the occurrence of power leakage issues as illustrated in Fig.~\ref{fig:AD_leakage_reason}, the sparse representations cannot accurately depict the actual angular power spectrum $A^2(\theta, \phi)$ of the HMIMO channel. 
In contrast, rather than sampling through the Dirichlet kernel as its angular-domain counterpart does, the sparse elements in wavenumber domain representation are integrals of the actual angular power spectrum $A^2(\theta, \phi)$ over the corresponding region $\Omega_l(\theta,\phi)$, which effectively isolates energy leakage from other spatial regions, leading to a more accurate depiction of the HMIMO channel as shown in Fig.~\ref{fig:AD_power_tail}(b).}

{\color{black}On the other hand,} the wavenumber-domain representation is independent of the number of holographic antenna elements and is even compatible with the continuous-aperture HMIMO designs. 
This is because the FH-based dictionary matrix is determined by the aperture size and the carrier wavelength. 
In particular, regarding the angular-domain representation in (\ref{AD_dic}), the column number of the dictionary matrix $\mathbf{ \Psi }$ is actually the number of antenna elements $N$. By contrast, the column number of the wavenumber-domain dictionary matrix is approximated by $L  \approx \pi \frac{L_x L_y}{\lambda^2}$~\cite[Fig. 5]{28.6}. To facilitate exposition, we present the ratio of the dimensionality between the angular-domain and wavenumber-domain dictionary matrices, i.e.,
\begin{equation}\label{ratio}
	\frac{N}{L } = \frac{{{\lambda ^2}}}{{\pi {\delta ^2}}} = 
	\begin{cases}
		\frac{4}{\pi} \approx 1.27  , & \delta = \frac{\lambda}{2}, \\
		\frac{16}{\pi} \approx 5.09 , & \delta = \frac{\lambda}{4}, \\
		\frac{64}{\pi} \approx 20.37 , & \delta = \frac{\lambda}{8}.
	\end{cases}
\end{equation}
Evidently, as the antenna elements are more densely packed, the dictionary in the wavenumber domain not only identifies the sparsity determined by the clusters in the propagation environment but also exhibits the reduced size of the sparsifying basis in contrast to its angular-domain counterpart. This facilitates more efficient CS algorithms, as will be elaborated on later.

\subsection{Channel Estimation Problem Formulation}
Let ${\mathbf{C}} = {\text{diag}}\left( {\mathbf{A}} \right){\mathbf{P}}{\text{diag}}\left( {\mathbf{M}} \right) \in {\mathbb{C}^{{N_f} \times N}}$ denote the measurement matrix, based upon which the channel estimation can be formulated as a CS problem
\begin{equation}
	\label{eq:WD_CS_formulation}
	\hat{\bf{h}}^f = \arg \min_{{\bf h}^f} 
	\left\{\left\| {\bf y} - {\bf C} {\bf \Psi}^f {\bf h}^f \right\|_2^2 + \lambda^p \left\| {\bf h}^f \right\|_p\right\},
\end{equation}
where $\lambda^p$ is the penalty parameter, and $\left\| \cdot \right\|_p$ represents the $p$-norm. The problem formulated in (\ref{eq:WD_CS_formulation}) aims to acquire an estimate of ${\mathbf{H}}$ by determining ${{\mathbf{h}}^f}$. Thus, it is required to determine both the indices and the amplitudes of the non-zero entries in the sparse channel vector ${{\mathbf{h}}^f}$.

%However, in the HMIMO channel estimation being considered, it is worthy mentioning that \(h_l^f\) follows a complex Gaussian distribution with a variance of ${\left( {\sigma _l^f} \right)^2}$. In the wavenumber-domain representation, the non-zero entry in the sparse channel vector, i.e., $h_l^f$, indicates a fragment of the channel gain determined by the cluster. It fails to encapsulate the angular power of a specific cluster and should not be misunderstood as the full channel gain. Instead, a collective contribution from other most correlated $h_l^f$ sampled are required to reconstruct the full angular power distribution corresponding to the cluster. By contrast, in traditional angular-domain channel estimation for mMIMO systems, identifying the indices and magnitudes of the non-zero entries in the sparse channel vector suffices to recover the associated significant paths. However, with the absence of `significant path' concept in the HMIMO context, we intend to directly reconstruct each cluster's angular power distribution to recover the full channel gain.

\section{Proposed Graph-Cut-based Swap Estimation (GCSE) Algorithm}
The clustered nature of the HMIMO sparse channel representations provides the underlying algorithm with additional prior structured information that can be exploited to improve the channel estimation accuracy further and reduce the computational complexity.
Specifically, consider specific entries in ${\bf h}^f$ (or ${\bf h}^a$), if we find a non-zero entry at the $l = ( l^{(0)}_x, l^{(0)}_y )$-th (or $n^\prime = ( {n^\prime}^{(0)}_x, {n^\prime}^{(0)}_y )$-th) wavenumber (or angular) domain index, then the corresponding entries in the neighboring $(l^{(0)}_x\pm 1, l^{(0)}_y\pm 1)$-th (or $({n^\prime}^{(0)}_x\pm 1, {n^\prime}^{(0)}_y\pm 1)$-th) wavenumber (or angular) domain indices are also likely to be non-zero.
This clustered sparse property can be encapsulated by tailored probabilistic models, such as a Markov random filed~\cite{11.18.cited.5,11.18.1, 11.18.6}.

{\color{black}
  Traditional methods like OMP are incapable of exploiting the clustered sparsity in the HMIMO channel estimation problem. Further, the brute-force searching in the one-by-one manner for the non-zero elements causes unacceptable computational complexity. 
  Therefore, we formulate the equivalent graph energy minimization problem by constructing the elliptic Markov random field (EMRF) model to exploit the clustered sparsity in the wavenumber domain, and propose a novel GCSE algorithm to solve this problem in a much more efficient manner.
  }

\subsection{Elliptic Markov Random Field}
\subsubsection{Motivation}
Numerous efforts have demonstrated that the naive Markov chain model shows great potential in capturing the clustered sparsity in the conventional ULA MIMO systems \cite{11.,cloudassisted, 12.}. 
However, regarding the HMIMO system considered in this work, the application of the chain-based model might be suboptimal due to its inability to capture the 2D spatial correlation involving both the elevation and azimuth angles, as well as and horizontal and vertical wavenumber-domain indices. Additionally, while the Markov random field (MRF) model is capable of capturing the spatial correlation in a 2D setting, its rectangular form may not be the best choice for accurately representing the elliptic shape of the wavenumber domain indices set, i.e., $\mathcal{L}$ shown in~(\ref{eq:xi_defination}). 
Consequently, there's a pronounced need for a bespoke solution adept at encapsulating the intricacies of the wavenumber domain.

\subsubsection{Topology Design}
A novel EMRF model is proposed to capture the clustered sparsity in the wavenumber domain. Specifically, we introduce {\color{black}an undirected graph} $\mathcal{G}\triangleq \left( \mathcal{V}, \mathcal{E} \right) $, where $\mathcal{V}$ and $ \mathcal{E}$ denote the vertex and edge sets of the EMRF, respectively.
Each vertex $v_l \in \mathcal{V}$ corresponds to the $l$-th wavenumber-domain index.
{\color{black}
  The undirected edge set $\mathcal{E}$ is defined as the indices pair of the neighboring vertex, i.e., }$\mathcal{E} \triangleq \left\{ \left\{ l, {l^{\prime}} \right\} \mid l, l^{\prime} \in \mathcal{L}, \left\| l - l^{\prime} \right\|_1 = 1 \right\}$.
Vertex $v_l, \forall l \in \mathcal{L}$ is defined as the binary variable with value pair $\left\{-1, 1\right\}$, where $v_l = 1$ denotes {\color{black}$h^f_l$} is non-zero, and $v_l = -1$ denotes $h^f_l$ is zero.
Therefore, ${\bf v} \triangleq \operatorname{vec} \left( \mathcal{V} \right)\in \left\{{-1, 1}\right\}^{L \times 1}$ can be formulated as the binary support vector of the wavenumber-domain sparse channel vector ${\bf h}^f$.

%in the field of hidden Markov models~{\color{red} xx}.

\subsubsection{Probabilistic Model}
The essence of capturing the clustered sparsity lies in the probabilistic modeling of the correlation between the neighboring vertices in EMRF.
Accordingly, we define $\eta_{l,l^\prime}\in \mathbb{R}^{+}, \forall \{l,l^\prime\}\in \mathcal{E}$ as the correlation controlling factor.
Therefore, the joint probability of the EMRF can be given by
\begin{equation}\small
  \label{eq:EMRF_joint_distribution}
  \begin{aligned}
    p\left({\bf v} ; \boldsymbol{\eta}\right) & = \exp \Big\{ \sum_{\left\{l, l^\prime\right\}\in \mathcal{E}} \eta_{l,l^\prime}\cdot v_l v_{l^\prime} + \sum_{l\in \mathcal{L} } \eta_l \cdot v_l - Z({\boldsymbol{\eta}}) \Big\} \\
    &\stackrel{(a)}{\propto}\exp \Big\{ \sum_{\left\{l, l^\prime\right\}\in \mathcal{E}} \eta_{l,l^\prime} \cdot v_l v_{l^\prime}+ \sum_{l \in \mathcal{L}} \eta_l \cdot v_l\Big\} \\
    &\stackrel{(b)}{\propto} \exp \Big\{ \sum_{\left\{l, l^\prime\right\}\in \mathcal{E}} \eta_{l,l^\prime} \cdot v_l v_{l^\prime} \Big\},
  \end{aligned}
\end{equation}
where we take the following assumptions:
\begin{itemize}
  \item[(a)] $Z\left(\boldsymbol{\eta}\right)$ is the normalization factor to ensure $\int_{\bf v} p\left( {\bf v}; \boldsymbol{\eta} \right) = 1$, which is independent of ${\bf v}$.
  \item[(b)] $\eta_l$ denotes the prior probability of the $l$-th vertex being non-zero, which is assumed to be fixed parameters.
\end{itemize}
Then, the corresponding observation model associated with the received signal presented in~(\ref{eq:transmit_model}) can be formulated as
\begin{equation}
  \label{eq:conditional_probability}
  p\left({\bf y} | {\bf h}^f\right) = \mathcal{CN} \left( {\bf y}; {\bf C} {\bf \Psi}^f {\bf h}^f, \sigma_{\rm noi}^2 {\bf I} \right),
\end{equation}
where $\sigma_{\rm noi}^2$ represents the noise variance.

%\begin{figure*}[htbp]
%  \centering
%  \includegraphics[width=1.\textwidth]{figure/visualization.pdf}
%  \caption{Visualization of the sparse channel vector ${\bf h}^f$ and the corresponding support vector ${\bf v}$ in the wavenumber domain.}
%\end{figure*}

\subsection{Maximum A Posteriori (MAP) Estimation}

%({\color{red} why need MAP estimation and the logic of proposing the MAP estimation and the so-called structured information})

Note that the joint distribution of $ \left\{ {\bf v}, {\bf h}^f, {\bf y} \right\}$ is proportional to the posterior distribution of $ \left\{ {\bf v}, {\bf h}^f \right\}$, yielding that
\begin{equation}
  \label{eq:posterior_distribution}
  p\left( {\bf v}, {\bf h}^f\mid {\bf y} \right) \propto p\left( {\bf y}, {\bf v}, {\bf h}^f \right).
\end{equation}
As a result, the MAP estimation of the sparse channel vector ${\bf h}^f$ can be equivalently formulated by maximizing the joint distribution, which can be further refined as
\begin{equation}
  \begin{aligned}
    p\left({\bf v}, {\bf h}^f, {\bf y}\right) &= p\left({\bf v}, {\bf h}^f\right)  p\left({\bf y} \mid {\bf v}, {\bf h}^f\right) \\
    &\stackrel{(a)}{=} p\left({\bf v}\right)  p\left({\bf h}^f \mid {\bf v}\right) p\left({\bf y} \mid {\bf h}^f\right),
  \end{aligned}
\end{equation}
where (a) holds due to assuming the independence of ${\bf v}$ and ${\bf h}^f$~\cite{11.}.
By substituting~(\ref{eq:EMRF_joint_distribution}) and~(\ref{eq:conditional_probability}) into the above equation, we obtain the joint distribution as follows:
\begin{equation}
  \vspace{-0.2em}
  % p\left({\bf v}, {\bf h}^f, {\bf y}\right) \propto \exp\left\{ \sum_{\left(l, l^\prime\right) \in \mathcal{E}} \eta_{l,l^\prime}\cdot v_l v_{l^\prime} + \sum_{l\in \mathcal{L}}\eta_l\cdot v_l \right\} \cdot \prod_{l\in \mathcal{L}} \exp\left\{ \log \left(p\left(h_l^f\mid v_l\right)\right) \right\} \cdot \exp \left\{ \frac{1}{2\sigma_{\rm noi}^2} ||{\bf y} - {\bf C} {\bf \Psi} {\bf h}^f || \right\}.
  p({\bf v}, {\bf h}^f, {\bf y}) \propto
  \exp\left\{ 
    \mathcal{P}_{P} + \mathcal{P}_{S} + \mathcal{P}_{L}
   \right\},
\end{equation}
where $\mathcal{P}_{P} $, $\mathcal{P}_{S}$, and $\mathcal{P}_{L}$ are explicitly given by
\begin{subequations}\small
  % \vspace{-0.4em}
  \label{eq:three_terms}
  \begin{align}
    &\mathcal{P}_{P} = \sum_{\left(l, l^\prime\right) \in \mathcal{E}} \eta_{l,l^\prime}\cdot v_l v_{l^\prime} + \sum_{l\in \mathcal{L}}\eta_l\cdot v_l, \\
    &\mathcal{P}_{S} = \sum_{l\in \mathcal{L}} \log \left(p\left(h_l^f\mid v_l\right)\right), \\
    % \vspace{-0.4em}
    &\mathcal{P}_{L} = - ||{\bf y} - {\bf C} {\bf \Psi}^f {\bf h}^f ||^2/({2\sigma_{\rm noi}^2}).
  \end{align}
\end{subequations}
\subsubsection{{Prior Information}}
Within the term $\mathcal{P}_{P}$ that contains the prior information, the design of the sum term $\sum_{\left\{l, l^\prime\right\} \in \mathcal{E}} \eta_{l,l^\prime}\cdot v_l v_{l^\prime}$ is inspired by the {Ising} model~\cite{11.18.6}.
By adjusting the weight $\eta_{l,l^\prime}$, we can control how likely the neighboring vertices are to have the same value.
Meanwhile, since prior probability of the vertex $p\left(v_l\right)$ is unavailable, the term $\sum_{l\in \mathcal{L}}\eta_l\cdot v_l$ can be set to a constant, which can be incorporated to the normalization factor $Z\left(\boldsymbol{\eta}\right)$.
\subsubsection{{Support Information}}
The hidden support vector ${\bf v}$ is the binary vector that indicates whether the corresponding entry in ${\bf h}^f$ is zero or not.
Therefore, the structured information inherent in wavenumber-domain sparse vector ${\bf h}^f$ can be efficiently captured by carefully designing the support vector transition probability
\begin{equation}
  \label{eq:support_transition_probability}
    p\left(h_l^f\mid v_l\right) = \mathcal{CN} \left(0, \left(\sigma_l^f\right)^2 \right)^{\mathbb{I} \left( v_l = 1 \right)} \delta \left( h_l^f \right)^{\mathbb{I} \left( v_l = -1 \right)},
%    &\approx \mathcal{CN} \left(0, \left(\sigma_l^f\right)^2 \right)^{\mathbb{I} \left( v_l = 1 \right)} \mathcal{CN} \left(0, \epsilon^2 \right)^{\mathbb{I} \left( v_l = -1 \right)},
\end{equation}
where $\mathbb{I} \left( \cdot \right)$ denotes the indicator function and $\delta \left( \cdot \right)$ denotes the Dirac delta function. $\delta \left( h_l^f \right)$ can be approximated by a complex Gaussian distribution with zero mean and variance $\epsilon^2$, i.e., $\delta \left( h_l\right)  \approx \mathcal{CN} \left(0, \epsilon^2 \right)$, where $\epsilon^2$ is determined by the average residual power over the zero points, i.e., $\epsilon^2 \triangleq \frac{||{\bf r}||^2}{|\{v_l|v_l=-1\}|}$.

\begin{algorithm}[t]
  % \vspace{2cm}
  \caption{GCSE Algorithm}
  \raggedright 
  \textbf{Initialize:} ${\bf h}^{f, (0)} {\rm =} {\bf 0}, {\bf v}^{(0)} {\rm =} \{-1\}^{L}$, {\color{black}residual ${\bf r}^{(0)} = {\bf 0}$}, $j = 0$.
  \vspace{-1.2em}
  \begin{algorithmic}[1]\label{alg:GCSE}
    \REQUIRE ${\bf C}$, ${\bf \Psi}^f$, ${\bf y}$ and pre-defined $\widetilde{K}$.
    \WHILE {$j\leq J_M$ or ${\color{black}||{\bf r}^{(j)}||} \geq \epsilon_r$} 
      \STATE {$j=j+1$}
      \STATE {Calculate residual ${\bf r}^{(j)} = {\bf y} - {\bf C} {\bf \Psi}^f {\bf h}^{f, (j-1)}$.}
      \STATE {Update support vector ${\bf v}^{(j)}$ by~(\ref{eq:alpha_beta_swap}).}
      \STATE {Update sparse channel estimate $\hat{\bf h}^f = {\bf h}^{f, (j)}$ by~(\ref{eq:residual_minimization}).}
    \ENDWHILE
    % \STATE $sum \leftarrow a + b$
    \RETURN $\hat{\bf h}^{f}$
  \end{algorithmic}
\end{algorithm}

\subsubsection{Likelihood Information}
The likelihood probability is derived from the conditional probability of the observation model in~(\ref{eq:conditional_probability}).
The maximization of (\ref{eq:three_terms}c) is equivalent to minimizing the Euclidean distance between the observation ${\bf y}$ and the estimated observation ${\bf C} {\bf \Psi}^f {\bf h}^f$.
In other words, $\mathcal{P}_{L}$ represents how likely estimated sparse channel ${\bf h}^f$ is to be aligned with observation ${\bf y}$.

\subsection{Alternating Update}

Given the facts in~(\ref*{eq:three_terms}), the maximum likelihood (ML) estimation shown in~(\ref{eq:WD_CS_formulation}) can be recast to
\begin{equation}
  \vspace{-0.2em}
  \label{eq:ML_estimation}
  \hat{\bf{h}}^f_{\rm ML} = \arg \max_{{\bf h}^f} \left\{ \mathcal{P}_{L} - \lambda^p ||{\bf h}^f||_p \right\},
\end{equation}
which lacks the prior information provided by support vector ${\bf v}$.
Therefore, we formulate the MAP estimation of sparse channel vector ${\bf h}^f$ as follows:
\begin{equation}
  \label{eq:MAP_formulation}
\resizebox{0.85\hsize}{!}{$
    \hat{\bf{h}}^f_{\rm MAP} = \arg \max_{{\bf h}^f, {\bf v}} \Big\{ \mathcal{P}_{L}- \lambda^p ||{\bf h}^f||_p 
     + \mathcal{P}_{P} + \mathcal{P}_{S} 
     \Big\}.
$}
\end{equation}
Given the presence of the support vector ${\bf v}$, the joint estimation of ${\bf h}^f$ and ${\bf v}$ is a combinatorial optimization problem, which is NP-hard \cite{cloudassisted}.
To tackle this challenge, we resort to the alternating optimization strategy. Specifically, in each iteration, we can alternatively update ${\bf h}^f$ and ${\bf v}$, by resolving the sub-problems formulated below
\begin{subequations}
  \label{eq:AO}
  \begin{align}
    &\hat{\bf v} = \arg \max_{{\bf v}} \left\{ \mathcal{P}_{P} + \mathcal{P}_{S} \left({\bf v} ;\bar{\bf h}^f\right) \right\},
     \\
      \hat{\bf h}^f &= \arg \min_{{\bf h}^f} \Big\{ -\mathcal{P}_{L} + \lambda^p ||{\bf h}^f||_p 
       -\mathcal{P}_{S} \left({\bf h}^f ; \bar{\bf v}\right) \Big\},
  \end{align}
\end{subequations}
where $\bar{\bf v}$ and $\bar{\bf h}^f$ denote the support vector and sparse channel vector obtained in the previous iteration, respectively.
% \hl{the following content is the way to address the problem formulated in (21)}

\begin{table}[t]
	\centering
		\caption{Edge weight definition for $\alpha$-$\beta$ swap expansion}
	\begin{tabular}{|c|c|c|}
		\hline
		Edge & Weight & For \\
		\hline
		\hline
		$t^\alpha_l$ & $E(v^{j-1}) = D_l(\alpha) + \sum_{l' \in N_l} V_{l,l'}(\alpha, v_l^{j-1})$ & $l \in \mathcal{L}$ \\
		\hline
		$t^\beta_l$ & $E(v^{j-1}) = D_l(\beta) + \sum_{l' \in N_l} V_{l,l'}(\beta, v_l^{j-1})$ & $l \in \mathcal{L}$ \\
		\hline
		${l,l'}$ & $V_{l,l'}(\alpha, \beta)$ & $\{l, l'\} \in \mathcal{E}$ \\
		\hline
	\end{tabular}
	\label{tab:your_label}
\end{table}

\subsubsection{{Sparse Channel Estimation}}
As observed in~(\ref{eq:AO}b), the sparse channel estimation problem is in the form of a standard residual minimization problem with an extra term, i.e., $\left\{ -\mathcal{P}_{S} \left({\bf h}^f ; \bar{\bf v}\right)\right\}$.
Therefore, we can rewrite the corresponding procedure as follows:
\begin{equation}
  \label{eq:residual_minimization}
  \resizebox{0.87\hsize}{!}{$
  % \begin{align}
    % {\bf r} &= {\bf y} - {\bf C} {\bf \Psi}^f {\bf h}^{f,\left(j-1\right)}, \\
    % {\bf t} &= \left({\bf C} {\bf \Psi}\right)^H {\bf r} + {\bf h}^{f, \left(j-1\right)},\\
    \hat{\bf h}^{f}  = \operatorname{Trim} \left\{ {\bf 0 }^{L} \left[:, \bar{v}_l = 1\right]= \left( ({\bf C} {\bf \Psi}^f) \left[:, \bar{v}_l = 1\right]\right)^\dagger {\bf y}; \widetilde{K} \right\},
    $}
  % \end{align}
\end{equation} 
where $\widetilde{K}$ is the pre-defined number of the non-zero elements in the sparse channel vector ${\bf h}^f$, $\operatorname{Trim} \left\{ \cdot; \widetilde{K} \right\}$ denotes the operation that keeps the $\widetilde{K}$ largest elements in the vector and sets the rest to zero. $(\cdot)^\dagger$ denotes the pseudo inverse.

\subsubsection{Support Vector Estimation}
% {\hl{miss some explanations?}}
% \begin{align}  \label{eq:graph_cut}
%     \hat{\bf v}  = \arg \max_{{\bf v}\in \left\{ \pm 1\right\}^{L} } \left\{ \mathcal{P}_{P} + \mathcal{P}_{S} \left({\bf v} ;{\bf t}\right) \right\},
% \end{align}
% where ${\bf t} = {\bf \Phi} ^H \left({\bf y} - {\bf \Phi } \bar{\bf h}^{f}\right) + \bar{\bf h}^{f}$. Then, (\ref{eq:graph_cut}) 
(\ref{eq:AO}a) can be equivalently formulated as a graph energy minimization problem, which can be solved by the graph-cut-based  algorithms~\cite{GraphCuts}.
\begin{align}
  \label{eq:energy_minimization}
    \hat{\bf v} &=\min_{{\bf v} \in \left\{ \pm 1 \right\}^{L} } E\left({{\bf v}; \bar{{\bf h}}^f}\right) \nonumber \\
    &= \min_{{\bf v} \in \left\{ \pm 1 \right\}^{L} } \Big\{ {\sum_{\left(l,l^\prime\right)\in \mathcal{E}}V_{l,l^\prime}\left( v_l, v_{l^\prime} \right) } + {\sum_{l\in \mathcal{L}} D_l\left(v_l\right)} \Big\},
\end{align}
where $V_{l,l^\prime}\left( v_l, v_{l^\prime} \right) = -\eta_{l,l^\prime} \cdot \left(v_l, v_{l^\prime} - 1\right)$ denotes the edge energy, $D_l\left(v_l\right) = -\log\left(p\left(\bar{h}_l^f|v_l\right)\right)$ for the vertex energy and $E\left({{\bf v}; \bar{{\bf h}}^f}\right)$ for the total graph energy.

\begin{figure*}[htbp]
% \vspace{1em}
  \centering
  \begin{minipage}{0.50\textwidth}
      \centering
      \includegraphics[width=\linewidth]{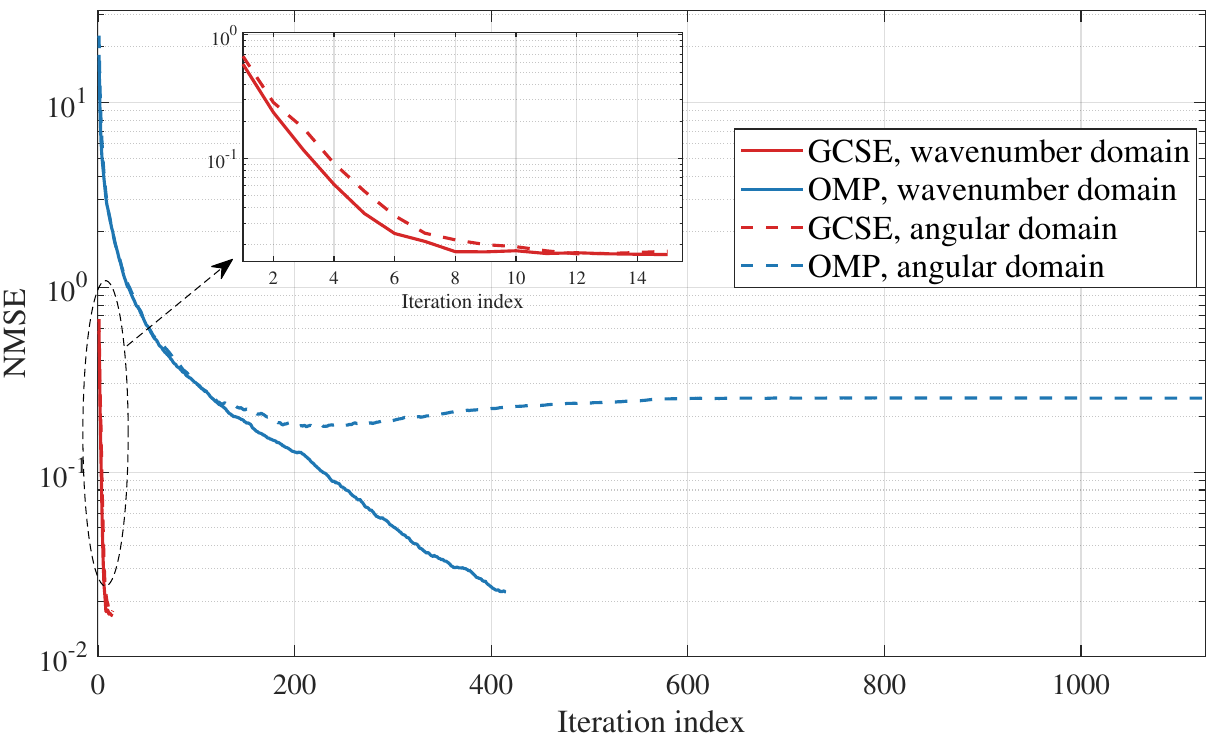}
      \caption{Convergence speed comparison.}
      \label{fig:convergence}
  \end{minipage}
  \hfill
  \begin{minipage}{0.48\textwidth}
      \centering
      \includegraphics[width=\linewidth]{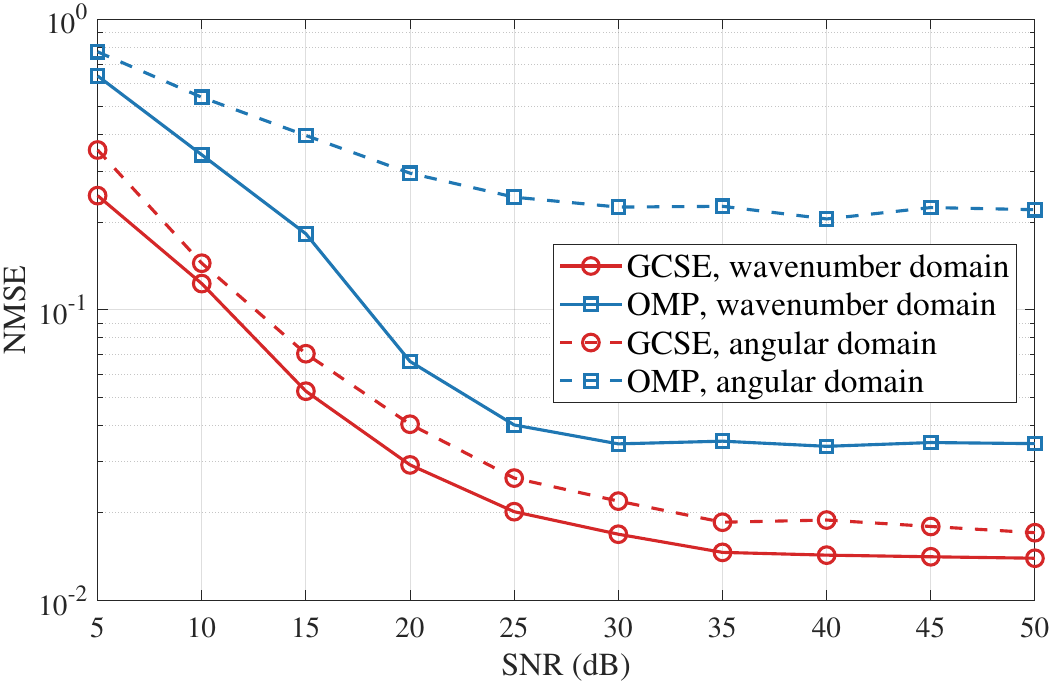}
      \caption{NMSE vs. SNR.}
      \label{fig:NMSE}
  \end{minipage}
  % \vspace{-0.03cm}
\end{figure*}

\subsection{$\alpha$-$\beta$ Swap Expansion}
As formulated in~(\ref{eq:energy_minimization}), the support estimation problem has been converted into a graph energy minimization problem. To achieve the optimum of~(\ref{eq:energy_minimization}), we define a new graph $\mathcal{G}^\prime$ based on earlier defined EMRF $\mathcal{G}$ by including the two label vertices, i.e., $\left\{\alpha = 1, \beta = -1\right\}$. Then, we introduce the edge weight shown in Table~\ref{tab:your_label}, where $t^\alpha_l$ and $t^\beta_l$ represent the edges connecting the vertex $v_l$ to the label vertices $\alpha$ and $\beta$, respectively.

In graph $\mathcal{G}^\prime$, we define the vertex $\alpha $ as the {\it source} terminal, while the vertex $\beta$ is the {\it sink} terminal. Then, in light of \cite[Corellary 4.5]{GraphCuts}, if we find the minimum cut $\mathcal{C}\left(\mathcal{G}^\prime;{\bf v}^{(j-1)}\right)$ through the maximum flow calculation (the computational complexity of which is only linearly proportional to the size of $\mathcal{G}^\prime$), we can arrive at the optimal solution of~(\ref{eq:energy_minimization}), i.e.,
\begin{equation}
  \label{eq:alpha_beta_swap}
v_l^{(j)} = v_l^{\mathcal{C}\left(\mathcal{G}^\prime;{\bf v}^{(j-1)}\right)} = \begin{cases}
  \alpha, & \text{if } t_l^\alpha \in \mathcal{C}\left(\mathcal{G}^\prime;{\bf v}^{(j-1)}\right) \\
  \beta, & \text{if } t_l^\beta \in \mathcal{C}\left(\mathcal{G}^\prime;{\bf v}^{(j-1)}\right)
\end{cases}.
\end{equation}
The overall GCSE algorithm is summarized in \textbf{Algorithm~\ref{alg:GCSE}}.

% \subsection{Benefits Analysis}
% \subsubsection{Convergence Speed}

% \subsubsection{Computational Complexity \& Signaling Overhead}
% \begin{table*}
%   \centering
%   \begin{tabular}{|c|c|c|c|c|}
%   \hline
%    & GCSE\_WD & OMP\_WD & GCSE\_AD & OMP\_AD \\
%   \hline
%   CPU time \& Iteration number & 10s, 15 iters & 1m12s, 415 iters & 29s, 15 iters & 13m12s, 1128 iters \\
%   \hline
%   Complexity Order & $O(M_{\text{RF}}I_s + M_K I_s[2])$ & $O(K_{\text{RF}}I_s + K^2 I_s[2])$ & $O(M_{\text{RF}}N + M_K N^2)$ & $O(K_{\text{RF}}N + K^2 N^2)$ \\
%   \hline
%   \end{tabular}
%   \caption{Your Table Caption Here}
%   \end{table*}

% \begin{table}
%   \centering
%   \begin{tabular}{|c|c|c|c|}
%   \hline
%    & CPU time \& Iteration number & Complexity Order \\
%   \hline
%   GCSE\_WD & 10s, 15 iters & $O(M_{\text{RF}}I_s + M_K I_s[2])$ \\
%   \hline
%   OMP\_WD & 1m12s, 415 iters & $O(K_{\text{RF}}I_s + K^2 I_s[2])$ \\
%   \hline
%   GCSE\_AD & 29s, 15 iters & $O(M_{\text{RF}}N + M_K N^2)$ \\
%   \hline
%   OMP\_AD & 13m12s, 1128 iters & $O(K_{\text{RF}}N + K^2 N^2)$ \\
%   \hline
%   \end{tabular}
%   \caption{Your Table Caption Here}
% \end{table}

\section{Simulation Results}

\subsection{Simulation Setup}
In this section, we conduct performance evaluations of the proposed GCSE algorithm for channel estimation in the HMIMO system. 
The $N_x = N_y = 129$ antenna elements are packed in a dense HMIMO UPA with antenna space $\delta = \lambda / 4$ and $N_f=1,000$ feeds at the BS.
The considered HMIMO system operates at 7~GHz with a narrowband system setup. The SNR is given by $||{\bf C} {\bf \Phi}^i {\bf h}^i ||_2^2 / \sigma_{\rm noi}^2,\ i\in \left\{a, f \right\}$ {\color{black}and the normalized mean square error (NMSE) is defined by ${||\hat{\bf H} - {\bf H}||_2^2 / || {\bf H}||_2^2}$.}
We compare the proposed GCSE algorithm with the traditional OMP algorithm in terms of both convergence speed and channel estimation precision.

\subsection{Performance Evaluation}
% As shown in Fig.~\ref{fig:convergence}, 
\subsubsection{Convergence Behavior}
{\color{black} 
As demonstrated in Fig.~\ref{fig:convergence}, the convergence speed of the proposed GCSE algorithm is significantly faster than that of the traditional OMP algorithm. This acceleration is attributed to the fact that, in each iteration, the GCSE algorithm updates multiple non-zero elements simultaneously, as opposed to the one-by-one searching approach employed in the traditional OMP algorithm.
Moreover, regardless of the algorithm employed, the convergence speed under wavenumber domain modeling is notably faster compared to traditional angular domain modeling. 
This efficiency stems from the fact that the dictionary matrix size in the wavenumber domain is solely related to the UPA aperture, rather than being proportional to the number of antenna elements, as is the case in the angular domain.
}

\subsubsection{Robustness}
{\color{black}
  Fig.~\ref{fig:NMSE} shows the NMSE performance of the proposed GCSE algorithm and the traditional OMP algorithm under both wavenumber domain and angular domain modeling.
  As can be seen, the proposed GCSE algorithm consistently outperforms the traditional OMP algorithm in terms of NMSE performance, regardless of the modeling domain and SNR. 
  Even in the low SNR regime, the GCSE algorithm still delivers robust performance by utilizing the prior information brought by the clustered structure, while the traditional OMP algorithm fails to converge with acceptable accuracy.}

\section{Conclusion}
In this paper, we have investigated the sparse channel estimation in HMIMO systems.
Given that conventional angular-domain CS techniques always grapple with power leakage issue that erodes the detection accuracy, we revisit the HMIMO channel estimation in the wavenumber domain. Specifically, a wavenumber-domain sparsifying basis is customized by harnessing the clustered sparsity inherent in HMIMO channels. Then, the channel estimation is recast to the CS recovery of the angular power spectrum of the HMIMO channel in the wavenumber domain. To efficiently solve this CS problem, a GCSE algorithm is proposed for the rapid retrieval of clustered non-zero entries within the EMRF model, which captures the structured priori for MAP estimation.
Simulation results demonstrate that wavenumber-domain modeling provides a representation exactly reflecting the physical channel propagation characteristics of HMIMO with a computational complexity independent of the number and density of antenna elements. 
Furthermore, our proposed GCSE algorithm reveals a remarkable improvement in convergence speed and robust performance, even in low SNR scenarios.

\vspace{1em}
\section*{ACKNOWLEDGMENT}
This work was supported by the National Key R\&D Program of China under grant 2022YFA1003901; in part by Beijing Natural Science Foundation under Grant 4222011, and in part by the BUPT Excellent Ph.D. Students Foundation under Grant CX2023145. (\textit{Corresponding authors: Ying Wang; Zhaocheng Wang.})
% This work is supported by the National Key R&D Program of China under grant, under Grant 2022YFA1003901.
\bibliographystyle{IEEEtran}
% \nocite{*}
% \bibColoredItems{black}{mMIMO} 
% \bibColoredItems{black}{28.6.} 
% aaa
\bibliography{ICC_Holo_CE}      
\end{document}